\documentclass[a4paper,showkeys,showpacs]{revtex4}
\usepackage{psfig}
\usepackage{graphicx}
\usepackage{url}
\usepackage[small, bf]{caption}
\begin{document}

\title{Coupled applications on distributed resources}

\author{P.~V.~Coveney}
\author{G.~De~Fabritiis}
\author{M.~J.~Harvey}

\affiliation{Centre for Computational Science, Department of Chemistry, 
University College London, 20 Gordon Street, London, WC1H 0AJ, UK.}

\author{S.~M.~Pickles}
\author{A.~R.~Porter}

\affiliation{Manchester Computing, University of Manchester, 
Oxford Road, Manchester, M13 9PL, UK.}

% Note; The OED gives 'modelling' and 'modeling' as equivalent spellings
% and I prefer 'modelling' :-) 

%\date{March, 2005}

%------------------------------------------------------------------

\begin{abstract}
Coupled models are set to become increasingly important in all aspects
of science and engineering as tools with which to study complex
systems in an integrated manner. Such coupled, hybrid simulations
typically communicate data between the component models of which they
are comprised relatively infrequently, and so a Grid is expected to
present an ideal architecture on which to run them.  In the present
paper, we describe a simple, flexible and extensible architecture for
a two-component hybrid molecular-continuum coupled model (hybrid MD). 
We  discuss its deployment on distributed resources and the extensions 
to the RealityGrid computational-steering system to handle coupled models. 
\end{abstract}

%------------------------------------------------------------------
\pacs{83.10RP, 47.11.Mn, 47.11.-j}
\keywords{Molecular dynamics, fluid dynamics, coupled models, 
Grid computing, computational steering, online visualization}
\maketitle

%------------------------------------------------------------------
\begin{section}{Introduction}

Multi-scale modelling of physical systems is often employed to enable
an accurate but computationally expensive model to be applied to the
regions of a system where it is essential while a less accurate model
can be used for the remainder.  This enables far larger systems and longer
timescales to be modelled than can be tackled using the accurate model
alone.

A multi-scale model may be constructed by coupling together different
simulation codes, each specializing in modelling a region of the
system on one particular scale.  Such a model then consists of a set
of two or more codes which are synchronized through the (repeated)
exchange of information as they execute. It could also  be
written as a single executable code but, for flexibility, scalability
and ultimately, performance reasons, we consider the case of a coupled
model constructed from independent components which are interfaced to
allow them to exchange information.  These components are then free to
be deployed in such a way as to make the best use of available
resources, something that is particularly important in a Grid
environment.

%Grid computing is distributed computing performed transparently across multiple
%administrative domains at arbitrary geographical locations [a].
%Steering refers to the interactive
%adjustment of an executing code based on access to the state of the simulation
%and its evolution in time. While steering may involve concurrent use of
%visualisation to view the state of a simulation, this is not necessary and a
%key feature of the RealityGrid steering system is that it separates steering
%from visualisation [8]. A central requirement for such steering is
%checkpointing -- the ability to write out the entire state of a system at a
%chosen time step, from which a number of important capabilities become
%possible.  Amongst these, we have developed "malleable checkpoint migration",
%in which checkpoints can be moved among the various grid resources and
%restarted on differing numbers of processors from those on which the
%computation and associated checkpoint were originally distributed. This opens
%the door to a form of "performance control", which in general is concerned with
%optimising the execution of a computation, a challenging task on a grid
%comprised of a dynamically varying set of resources and rendered more complex
%for coupled applications.  
%The approach described in this paper extends the
%original RealityGrid steering architecture to accommodate coupled models and
%their steering, including performance control by malleable checkpointing and
%migration of the individual components.

Heart modelling provides a good example of multiscale
complexity.  Within the Integrative Biology (IB,
\url{http://www.integrativebiology.ac.uk}) project the use of coupled
models is being investigated in the area of heart disease and cancer
modeling \cite{gavaghan05, hey05}.  The IB project aims to develop an
integrated approach to multiscale computational modeling of these
diseases.  Currently, the results of a heart cell model are integrated
into an electro-mechanical finite element discretization of the heart
muscular fibers (\url{http://www.physiome.org}). In the future, the
molecular level description of ion channels and drugs interfering with
them could be integrated within a systems biology model of the heart cell
which simulates the system response to changes in parameters related
to these ion channels. Although  we are still far from a
complete integration of this type, coupling between the heart cell
model and a finite element discretization of the heart is routinely
performed to study the activation potential and possible routes to
the development of arrhythmias and defibrillation~\cite{rodriguez05}.

The complexity of coupling multiple descriptions implies that, in the
long term, coupled models may be better handled by a series of coupled
codes rather than a single monolithic application which will be
difficult to develop, maintain and deploy. One step in this hierarchy
of scales which addresses the coupling of multiple descriptions of
matter is what we call ``hybrid MD''.  This approach is based on a
multiscale hybrid scheme, in which two or more contiguous sub-domains
are dynamically coupled together, the simplest case corresponding to
one sub-domain being described by molecular dynamics, the other by
continuum fluid dynamics \cite{buscalioni03, buscalioni05,
buscalioniPhilTrans05,defabritiis06}. The motivation for this hybrid
approach to the description of fluids is that it enables us to tackle
problems that are not addressable by any one single technique; such
problems include situations in which hydrodynamic effects influence
molecular interactions and {\em vice versa} at interfaces, lipid bilayer
membranes and within individual macromolecules or assemblies of
them. Since the computational overhead of coupling the molecular and
continuum regions is very low, we can use this approach to perform
coupled simulations using small molecular dynamics (MD) domains which are free of finite
size effects, whilst reducing the computational cost compared with
fully atomistic simulations.

Over the course of the RealityGrid project
(\url{http://www.realitygrid.org}), the introduction of launching,
monitoring and steering functionalities to existing scientific codes
has proved
successful in and out a Grid environment
~\cite{ReG_steering03,ReG_PhilTransSteering04}. 
 The prevalent
software model for grid computing is that of a service-oriented architecture
(SOA). Software components in a SOA environment comprise loosely coupled
interoperable application services, implemented as Web services, accessible via
SOAP operations whose capabilities are described by Web service description
language documents. The SOA methodology underpins the software described here.
Indeed, the developments described in the present paper have evolved from the RealityGrid
project.
The process
of ``computational steering'' refers to the interaction of a scientist
with his/her running application.  At its simplest, this may consist
of monitoring the values of a few key variables to check the progress
of the application and possibly altering the values of parameters of
the simulation.
  More advanced uses may involve the
 scientist altering variables that control algorithms within the code
  and ultimately  the
 development of complex workflows.  
Such use enables  scientists to further develop and apply their
intuition to enhance the investigation of the system under study.
On-line visualization of the system is a powerful aid in this process
since it can provide the user with visual feedback on the consequences
of his/her steering activity.  The implementation of computational
steering for coupled models presents some challenges which require 
extension of the RealityGrid system to handle coupled codes.
Although these extensions were tested on hybrid MD, they are in fact
applicable to any coupled model.
 
The remainder of the paper is laid out as follows. In
section~\ref{sec:HMD} we discuss the hybrid MD  coupled model. 
We describe the architecture of this system,
the communication scheme used and the way in which synchronization of
the components is achieved. Following this, the use of the RealityGrid
framework in deploying and managing hybrid MD is described in
section~\ref{sec:deployment}. In section~\ref{sec:multi_components} we
describe the way in which we combine information from each component
 to produce a description of the coupled model as a
whole and thus allow a steering client to treat it as a single,
steerable application.  
\end{section}

%------------------------------------------------------------------

\begin{section}{Hybrid molecular dynamics}
\label{sec:HMD}

Hybrid molecular dynamics (hybrid MD) is a coupled model formed by two distinct
descriptions of matter. At the microscopic level, the system is
modelled by classical molecular dynamics, while the mesoscale is
described by a finite volume discretization of the equations of
fluctuating hydrodynamics~\cite{defabritiis06}. These two descriptions
resolve different spatial domains and interact through a `hybrid'
interface where the hydrodynamic variables are coarse/fine-grained and
data is transferred between models (for a detailed description
see~\cite{defabritiis06}).

\begin{subsection}{Model structure}
% Figure 1
% Figure 1
\begin{figure}[]
\centerline{\psfig{figure=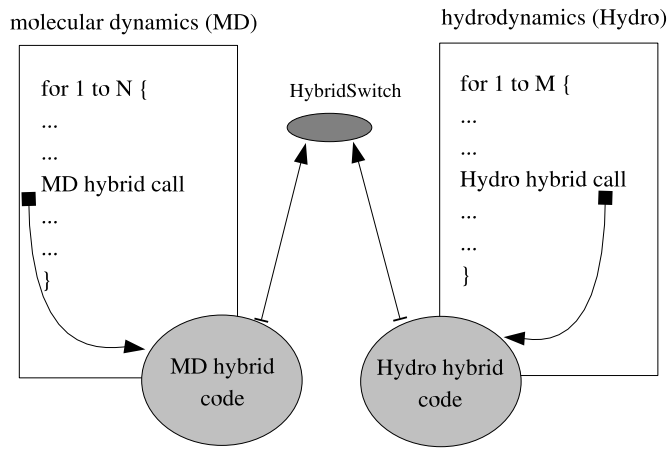,width=10.0cm}}
\caption{The implementation of the hybrid MD code requires minimum
perturbation of the original molecular dynamics (MD) and hydrodynamics
codes (Hydro).  The exchange and processing of information to
transform between physical descriptions is performed by opening a
channel in the main loop of each of the codes by calling a subroutine
in our `hybrid' library.  The two simulation components exchange data
by making calls to a library which performs the communications
operations via a third service (HybridSwitch).  These calls are placed
in the main loop of each code.  }
\label{fig:codearch}
\end{figure}

We have developed a hybrid MD scheme using two independent simulation
codes implementing, respectively, molecular dynamics (MD) and a finite
volume discretisation of the equations of fluctuating hydrodynamics
(Hydro)~\cite{defabritiis06}. These two models are heterogeneous from
a physical point of view but are also very different programmatically
because MD is a particle code, while Hydro is mesh-based. These
considerations and the need to have flexible and interchangeable
components for generalization to multiple components led us to create
a coupled architecture with two independent programs which exchange
data, rather than a single monolithic code.

This component-based approach has significant advantages. Not least of
these is that each individual component can be developed, tested and
used independently as a standalone code.  From a performance point of
view, the scheme is scalable under extension to multiple  (three, four,
{\it etc}) components as is required when adding extra levels of
description ({\it e.g.}  a quantum mechanical one) to the physical
model and possibly running each component on a different computer.

The architectural diagram illustrating how the two application
components interact is shown in figure~\ref{fig:codearch}.  The main
program of each application is modified in order to accommodate a call
to a subroutine from the `hybrid' library which takes care of
handling, preparing and communicating the data required by the hybrid
coupling. This `hybrid' code is rather specific and needs access to
the data structures of the programs (hydrodynamics or MD). In our case
we have hidden the code behind a simple interface which provides the
functionality needed. As a result, the `hybrid' code has to be written
only once and another molecular dynamics ({\it e.g.} another serial or
parallel) code could be deployed within the hybrid MD coupled model by
providing the same high-level interface. Of course, building this
interface requires dealing with the communication involved in
gathering all the information from each processor.  However, we do not
expect this to affect the performance of the codes because the hybrid
library is computationally light compared to the MD engine.

\end{subsection}

\begin{subsection}{Component synchronisation and data exchange}
\label{sec:synch}

A key issue for coupled models is the synchronisation of the various
components.  This is typically determined by the nature of the
information which the components must exchange with one another and
the stage during the calculation cycle at which this exchange occurs.
In the models we are dealing with, each component tackles a separate
physical region of the system and the data exchanged between them
provides the boundary conditions for each region.  The frequency of
this exchange is model dependent, {\it e.g.} the timestep of a
molecular dynamics  simulation might be one hundredth of that of
an associated continuum simulation.  In this case, the two components
might exchange data after every step of the continuum simulation (and
thus after every 100 steps of the MD simulation).  For the models we
consider here, each component is arranged to be at the same point in
(simulated) time whenever a data exchange takes place.

Although several specialised coupling frameworks exist, particularly
in the field of atmospheric and ocean
modelling~\cite{CCSM3,CISM_coupling,prism_coupling,OASIS3,OASIS4}, we
have chosen to use a very simple solution since our current models
contain only two components and do not use any particular features of
such frameworks.  To enforce synchronisation and permit data exchange,
we have constructed a communication ``switch'' service that implements
a store and forward message-passing scheme (here called HybridSwitch).
(Although this suffices for our current requirements, we mention in passing that other tools have
also been considered such as the Systems Biology Workbench
(SBW)~\cite{SBW}.)  The interface to this switch service consists
simply of a non-blocking {\em put} operation and a blocking {\em get}
operation.  It is the use of the latter that enforces inter-component
synchronisation.  In order to avoid deadlock in such a scheme, each
component must execute its `puts' and `gets' in a way that is
appropriate to the blocking behaviour of the communication.  Essentially,
this corresponds to executing the non-blocking
`puts' before the `gets' (see~\cite{gcf_coupling}).

Our experience within the RealityGrid project has consistently shown
that establishing some sort of socket-based connection between
applications running on different machines is plagued with problems.
It is not unusual for machines that are explicitly stated to be
`Grid-enabled' to have no facility for direct network connections to
the processors on which jobs are run (for instance, the core
`computational' nodes on the UK National Grid Service).  In the worst
case, the nodes on which a job runs might not have any internet
connectivity.  More common is the case where Network Address
Translation or tunnelling (via a head node) is used so that `back-end'
nodes can connect out to the Internet but do not themselves have
unique IP addresses, thus preventing incoming connections to software
running on these nodes, even if local firewall rules permit.
Introducing the HybridSwitch as a third party in order to mediate
communication between components alleviates some of these
difficulties.  In particular, any connection is always initiated by a
model component connecting {\em to} the HybridSwitch.  Therefore,
provided that any firewall around the machine running the HybridSwitch
is configured to allow incoming connections and that any firewalls on
the machines hosting components allow outgoing connections, the
communication will at least be free of firewall-related problems.

In our implementation, components connect to and communicate with the
HybridSwitch service, creating a communications network with a star
topology. This topology is perfectly adequate for the loose coupling
of hybrid MD.  For other types of applications, multiple HybridSwitches
could be deployed on different machines in order to share the network
load.  Each component is assigned a unique identifier
(hostname:port:jobid) which is used for message routing
purposes. Although indirecting messages through the HybridSwitch
imposes a latency penalty, there are several compensating benefits:
\begin{enumerate}
\item Loose coupling of components: Because messages are indirected,
an individual component need not know the  location of its
peers.
\item Transience of components: Because the HybridSwitch buffers undelivered
messages, components may be disconnected and reconnected without
destabilising the coupled application. It should be noted that this
may impose a performance penalty on any peer components performing a
blocking wait for messages from the disconnected component. When
combined with performance monitoring information, the ability to
seamlessly disconnect components would facilitate full performance
control~\cite{PerCo} through migration of components across resources.
\item Firewall-friendliness: This communication scheme is firewall-friendly 
because any connection is always initiated by the model
component and made to the HybridSwitch. Provided the HybridSwitch is located on a
resource with a suitable incoming-connection firewall policy,
individual components located on firewalled machines may still
intercommunicate.
\item Diagnostics:  The HybridSwitch provides a convenient, single point at
which any problems with the communications may be readily
investigated.
\end{enumerate}

\end{subsection}

\begin{subsection}{A Couette flow simulation}
\label{sec:couette}

We set up a simple numerical experiment in order to deploy and test
hybrid MD over distributed resources. The system comprises a
three-dimensional domain of $11$ physical cells of which the $3$ in
the centre are resolved with molecular dynamics (with the two external
ones providing a buffer to the MD system) while the other cells are
described by continuum fluctuating hydrodynamics~\cite{defabritiis06}
(figure \ref{fig:shot}).  The continuum model is terminated by no-slip
boundary conditions: the right wall does not move, while the left wall
moves at velocity $2$ {\AA}/ps.  The velocity profile produces a
momentum transfer across the cells which is transferred from the
continuum (cells 9-15) to the MD region (cells 7-9) and again to the
continuum (cells 1-6).  For this simple case the analytical solution
is known and, at the stationary state, is a linear decay of the
velocity profile (figure \ref{fig:shot}).

\end{subsection}

\end{section}

%----------------------------------------------------------------------

\begin{section}{Deploying a coupled model within the RealityGrid 
framework}
\label{sec:deployment}

The RealityGrid framework~\cite{ReG_PhilTrans_TeraGyroid} can be used to
deploy, steer and otherwise control the components of the hybrid MD coupled
model, including any associated visualization components.  Here we
describe the deployment and control of hybrid MD simulating 
a Couette flow (see section  \ref{sec:couette}) distributed over a
number of different machines. The HybridSwitch service runs 
on an independent machine listening on a given port. 

\begin{subsection}{Deployment}
\label{sec:Launch}

% Figure 2
% Figure 2
\begin{figure*}[]
\centerline{\includegraphics[width=15.0cm]{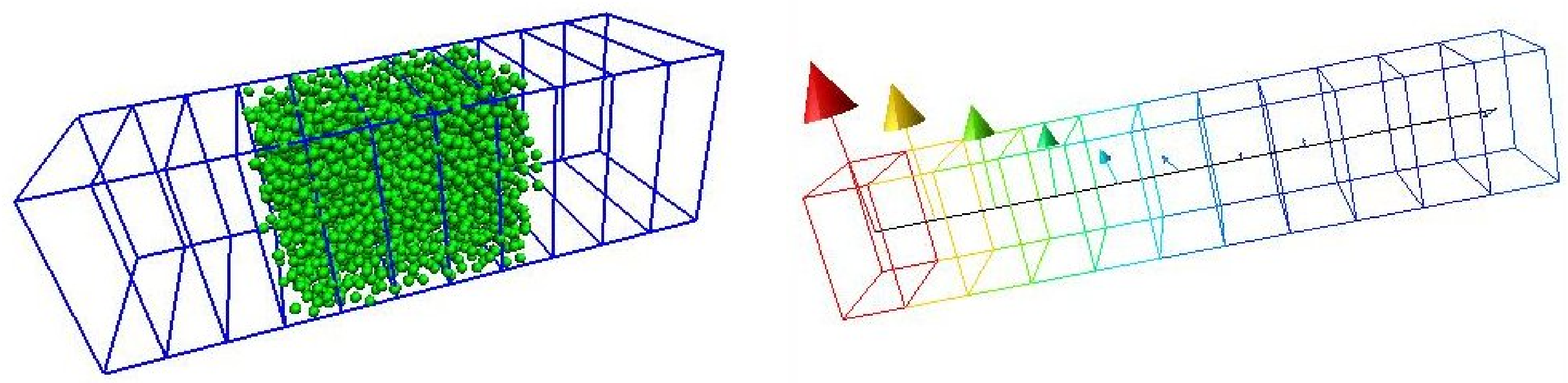}}
\caption{A picture of the output from the hybrid MD coupled model running on a
prototype system consisting of a box of atoms within a large
continuum domain.  On the left is the output of the visualization connected to the MD component
 showing the  system of atoms and the finite volume discretization for the fluctuating hydrodynamics. 
On the right is shown the velocity field during the Couette simulation from the hydrodynamics
component.  The visualization applications are themselves separate
components in the architecture.
Online visualization of the molecular region uses a
RealityGrid-modified version of VMD~\cite{HUMP96}.  An
AVS/Express-based visualization~\cite{reg_avs} is used to monitor the continuum
region. 
}
\label{fig:shot}
\end{figure*}

We use the RealityGrid launching client running on the user
workstation to deploy each component of the model onto suitable
resources.  As shown in the left of figure~\ref{fig:launch_gui}, the
user first selects which component of the coupled model to deploy (MD
or Hydro), any associated input files and the resource on which to run
it from drop-down lists (middle of figure~\ref{fig:launch_gui}).  If a
component is parallel then the user can also specify the number of
processors to use. Because of the coupling architecture provided by
the HybridSwitch, each component need only know the exact location and
port of the switch. Once the input files for all of the components
comprising the model have been entered, the launcher proceeds to start
each of them and their associated web services (right of
figure~\ref{fig:launch_gui} and section~\ref{sec:multi_components}).
Either secure shell (ssh) or Globus~\cite{globus} (according to user
preference and machine accessibility) is used to begin the execution
of each component on the remote resource chosen for it by the user.

During the launching process, the details of each component and the
coupled model as a whole are collected.  This information is then
published in a central registry allowing the user (or a collaborator)
to subsequently examine what jobs are running and possibly attach a
steering client to them.

% Figure 3
% Figure 3
\begin{figure*}[t!]
\centerline{{\includegraphics[width=12.0cm]{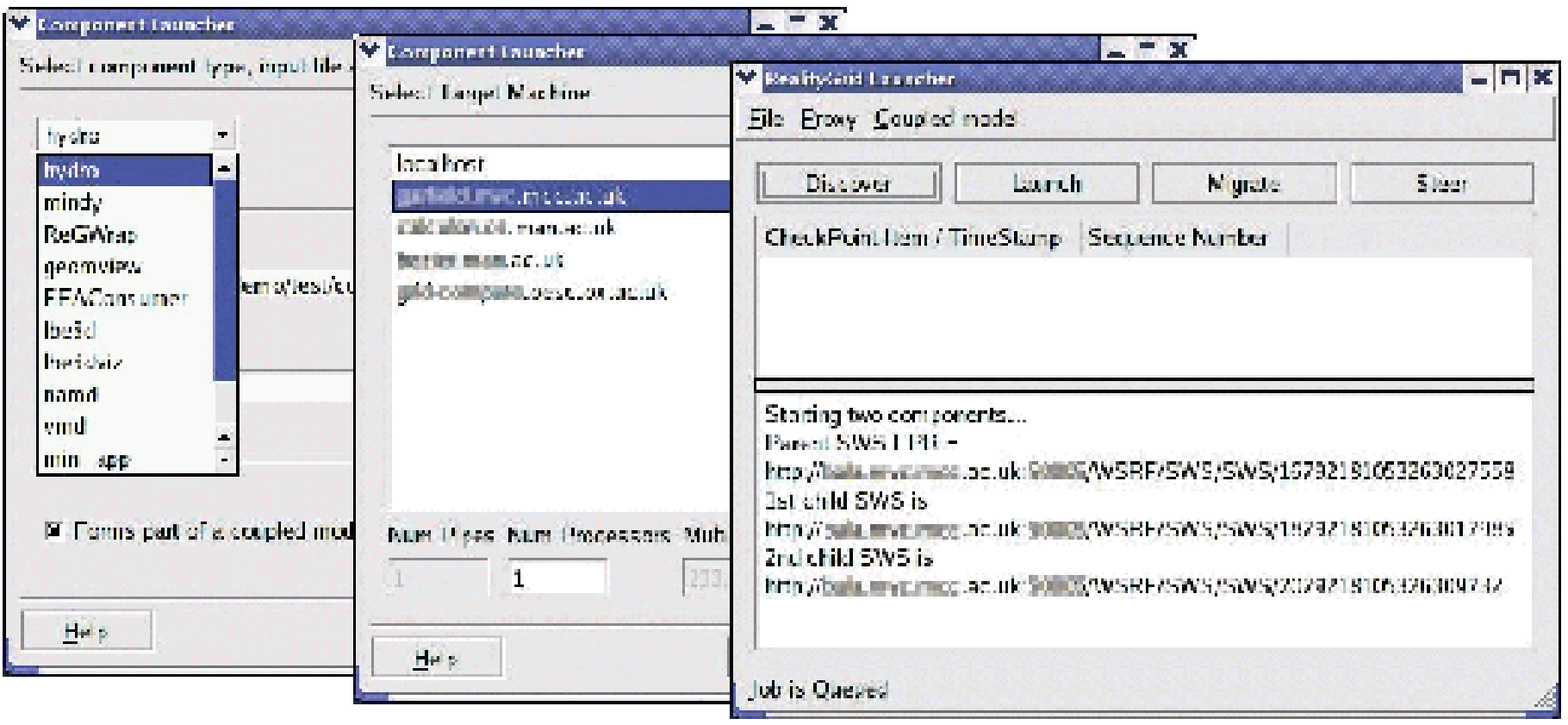}}
}
\caption{Snapshots of the RealityGrid launcher being used to deploy
the hybrid MD coupled model.}
\label{fig:launch_gui}
\end{figure*}

The launching client is also employed when the user wishes to deploy an
on-line visualization in order to monitor the state of a particular
component in more detail. In this case, the launcher interrogates the
registry and provides a list of the components that are currently
running.  The user selects the one to be used as a data source for the
new visualization and then proceeds as for a simulation component.
The creation of the (socket) connection between the visualization and
simulation components is mediated by the underlying web services.

\end{subsection}

%--------------------

\begin{subsection}{Performance Control}
\label{sec:Perf}

% Figure 4
% Figure 4
\begin{figure*}[t]
\centerline{\psfig{figure=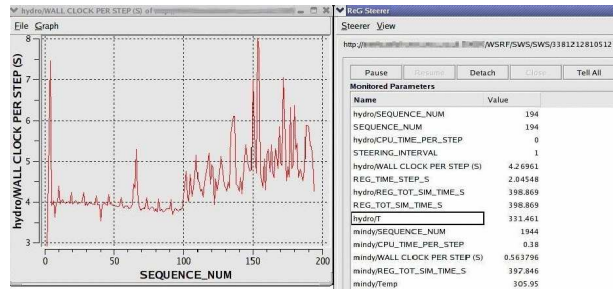,width=8.0cm}}
\caption{An illustration of the potential for monitoring the
performance of components within the hybrid MD coupled model using the RealityGrid
steering software.  The screenshot shows the wall-clock time per step
(as measured by the steering library) for the hydrodynamics application before and
during (beginning at timestep 100) the injection of an artificial
load to the machine on which it is running.}
\label{fig:load_monitor}
\end{figure*}
Once a coupled model has been made steerable, it becomes possible to
implement more advanced functionality.  We have previously created a
framework for job migration for single-component models using the
steering functionality and its support for
check-pointing~\cite{ReG_PhilTrans_TeraGyroid}. Job migration underpins
the ability to dynamically manage the performance of a running
application (\textit{e.g.}\ by migrating the job to  either a less
heavily-loaded machine  or to a faster
machine that was not available when the job was initially launched),
whether manually or automatically.

With the move to distributed, multi-component models, the need for
performance control becomes more pressing and the task more complex.
In general, each component of a coupled model will have different
performance characteristics and therefore the resources that must be
allocated to each to maximise the performance of the model as a whole
are not straightforward to determine.  Optimising the performance of
such a model therefore requires the determination of the rate-limiting
component and the subsequent migration of it to a more
powerful/suitable resource, if available. In the case of hybrid MD, the
perfomance on the specific machine can be determined on-the-fly by the
RealityGrid monitoring capabilities. In figure~\ref{fig:load_monitor},
the elapsed time per iteration of the Hydro component is monitored
while an artificial load is applied to the machine.  The user may then
decide to checkpoint the codes using the RealityGrid checkpoint
command and then restart it somewhere else. This provides a very first
simple and manual performance control over the coupled model as a
whole. In principle, the RealityGrid system and the HybridSwitch also
permit individual components to be stopped and restarted without
the loss of synchronization. However, hybrid MD does not support this feature
as yet.

Our existing job migration system is based upon the ability to
instruct a job to create a checkpoint and then record the details of
that checkpoint (location and names of files, snapshot of model
parameters \textit{etc.}) with a web
service~\cite{ReG_PhilTrans_TeraGyroid}.  For a coupled model a job
comprises two coupled applications, therefore the instruction to
create a checkpoint must be carried out by all components at the
equivalent point in {\em simulated} time.  The metadata for each
component's checkpoint must then be gathered together and registered
with the web service, as before.  Using this information it is then
possible to restart the coupled model as a whole on any set of
available resources after having first copied the necessary sets of
files to each machine.  We are currently working to implement this
functionality and hope to extend it so that migration of individual
components may be achieved without stopping and restarting the other
components of the model.

\end{subsection}

\end{section}

%------------------------------------------------------------------

\begin{section}{Representing multiple components as a single application}
\label{sec:multi_components}

% Figure 5
% Figure 5
\begin{figure*}[!t]
\centerline{\includegraphics[width=8.0cm]{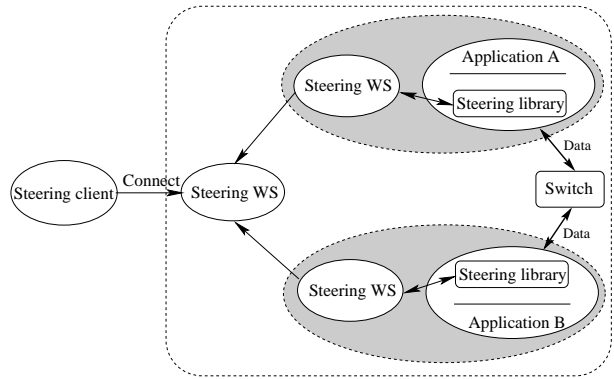}}
\caption{The architecture of a steered coupled model consisting of two
components with a separate ``switch'' (see section~\ref{sec:synch}) to
enable data transfer between them.  The coupled model appears as a
single component to the steering client.}
\label{fig:steer_arch}
\end{figure*}

The RealityGrid project has developed a computational steering API and
associated library~\cite{ReG_PhilTransSteering04, sve_site}, designed
to facilitate the instrumentation of an application code for
steering. Given that a large number of applications within the project
and an increasing number from outside it are instrumented with this
library, we aimed to design a system for steering coupled models that
would require no API changes.

Since a coupled model ultimately represents a single physical system,
it is more intuitive and thus highly desirable for a user wishing to
steer it to be able to treat it as such, rather than having to
consider its constituent components.  In addition, this approach has
the advantage that existing steering clients, originally designed to
connect to one or more independent components, can be used unchanged
to steer a coupled model consisting of a number of components.
In light of these requirements, the architecture of our system for
steering coupled models is of the form shown in
figure~\ref{fig:steer_arch}.  As in the standard RealityGrid steering
system~\cite{ReG_PhilTransSteering04}, each individual component is
represented by a Steering Web Service (SWS) --- a Web-Service Resource
having both state and lifetime.  The Web Services Resource Framework
(WSRF)~\cite{WSRF} provides standard interfaces for managing these
properties and we have implemented the SWS using the WSRF::Lite Perl
package~\cite{WSRF-Lite}.
In order for the coupled model to present a single interface to a
steering client, we introduce a third SWS as shown in the figure.
This SWS has responsibility for taking the information available to it
from its children (the SWSs representing each component) and
presenting it to the client in a suitable form.  Thus, the vast
majority of the extensions to the RealityGrid steering system to cope
with coupled models are restricted to the SWS. 

Figure~\ref{fig:steer_arch} shows a coupled model (e.g. hybrid MD) with only two
components, although our approach generalises to more complex
configurations, possibly requiring a deeper tree of SWSs or SWSs with
more than two children (both of which are supported).  In order for a
coupled model consisting of two or more components to present a single
steering interface, the various aspects of each component must be
combined in some way.  In our architecture, this is performed by any
SWS that has SWSs as its children.  Thus, in
figure~\ref{fig:steer_arch}, the left-most SWS is responsible for
making available the results of combining the information from the
two SWSs which communicate directly with the two components making up
the application.

Within the RealityGrid steering system, each steerable component is 
characterised by
certain metadata that is created by the steering library at runtime
using the information supplied by the application.  This metadata
consists of:
\begin{enumerate}
\item Steering commands (the set of steering commands supported by the
component like start, pause, stop, checkpoint, etc): 
Since steering commands are generic to a steerable
application, in order for a parent to support a command, it is
necessary and sufficient that all of its children support it.  Thus,
the set of steering commands supported by a parent may be generated by
ANDing the sets of commands supported by each of its children.
\item Monitored parameters (those parameters of the component that a
steering client can observe but not modify): The monitored parameters
of a component are specific to it since they are a part of the state
of that component. Although these parameters may represent the same
physical quantity, they provide useful information on the behaviour of
the two separate components and the correctness or otherwise of the
coupling between them.  Consequently, the childrens' sets of monitored
parameters are simply combined to form the set of monitored parameters
of the parent (albeit with tags added to their labels to ensure they
remain distinct).
\item IO channels and checkpoint types (IO channels support general
data IO whereas checkpoint types are specifically designed to support
control of check-pointing and restarting a component): As with
monitored parameters, the IO channels associated with each component
remain distinct.  These may be used to connect suitable visualization
packages to one or more of the components within the application and
thus need to retain their identity.  In contrast, check-pointing is a
more complex issue.  Provided that all of the components forming the
coupled model have registered at least one checkpoint type then, in
principle, it is possible to checkpoint the whole
application. However, for hybrid MD, it is important that each of the components
create their respective checkpoints at an equivalent point in
simulated time.
\item Steerable parameters (those parameters of the component whose
values can be altered by a steering client): Steerable parameters are
the most challenging part of a component's metadata to combine.  This
is because steerable parameters belonging to different children may
need to be represented by a single steerable parameter in the parent.
\end{enumerate}

The case of checkpoints and steerable parameters is better understood
with a specific  example from hybrid MD.  
For instance, MD and Hydro have
a monitored parameter representing the current local temperature which
fluctuates around a mean value given by the imposed thermostat
temperature.  Each component has a steerable parameter controlling the
{\em target} temperature which must be the same for both components.
(These component-specific parameters may well have different names,
{\it e.g.}\ \texttt{T\_target} and \texttt{desiredTemp}.)
% Figure 6
% Figure 6
\begin{figure*}[!t]
\begin{center}
\begin{verbatim}
    <SWS:Coupling_config>
      <Global_param_list>
        <Global_param name="targetTemp">
          <Child_param id="http://machine.address:10000/SWS/service/8312105489098098" 
                       label="8312105489098098/T_target"/>
          <Child_param id="http://machine.address:10000/SWS/service/8312105489023033" 
                       label="8312105489023033/desiredTemp"/>
        </Global_param>
      </Global_param_list>
    </SWS:Coupling_config>
\end{verbatim}
\end{center}
\caption{An example of metadata describing a single global parameter
for a two-component system.  The new, global parameter, is given the
name \texttt{targetTemp} and represents the parameter
\texttt{T\_target} from one child and \texttt{desiredTemp} from the
other.  The \texttt{id} attribute of each \texttt{Child\_param}
element identifies which child the original parameter belongs to.}
\label{fig:global_meta_data}
\end{figure*}

The RealityGrid steering system handles this case by defining such
parameters as being {\em global} within the scope of the parent. Any
change that a user makes to such a parameter must be passed down to
the components in such a way that the change occurs in synchronized
fashion, {\it i.e.} at equivalent points in simulated time. 
It is up to the developer of the coupled model to decide which parameters 
belong to a single description and which are intrinsically coupled.

A fragment of metadata to describe such global parameters is shown in
figure~\ref{fig:global_meta_data} where a new, global parameter
\texttt{targetTemp} is defined as being equivalent to one child's
\texttt{T\_target} parameter and the other child's \texttt{desiredTemp}
parameter.  This metadata is supplied to the parent SWS which then uses it to
perform the necessary processing of its childrens' parameters.  (Since all
messaging and metadata within the RealityGrid system uses XML, it is well
suited to processing in this way.) Once a parent SWS has used the supplied
metadata to process its children's steered parameters, the set of steered
parameters that it will supply to any attached steering client will contain the
\texttt{targetTemp} parameter in place of both \texttt{T\_target} and
\texttt{desiredTemp}.  Thus the steering client will have a single steerable
parameter to control the target temperature in both of the components.
Currently, our system assumes that the parameters making up a global parameter
are all in the same units although any necessary scaling factors could, in
principle, be included in the metadata.

When the user steers a global steered parameter, it is the parent
SWS's responsibility to ensure that the change to each of the
constituent parameters occurs at the same point in simulated time in
each component.  In order to achieve this, the parent SWS must have
knowledge of the current simulated time of each of its child
components.  This is achieved by requiring each component of a coupled
model to register a special steered parameter controlling the value of
the time step.  Given that the steering library can monitor the number
of timesteps completed, it can then also calculate the total amount of
time that has been simulated and make this available (as a monitored
parameter).  Given knowledge of each child component's time-step and
current simulated time, the parent SWS can calculate a suitable future
simulated time, $t_{target}$, at which to make the change to the
constituents of the global parameter.  It then inserts this
information into the control message (describing the change in
parameter value) being sent to each child.  The steering library has
been extended to make use of this information --- a message tagged in
this way is stored and only acted upon once the specified $t_{target}$
time has been exceeded by the component.

The same approach is used when check-pointing. A
$t_{target}$ is generated and included with the command to checkpoint
which is sent to each component.  It is worth noting that this
extension to the steering library provides most of the functionality
that is necessary to enable a user to specify when certain
steering actions are to be applied to a component.  For instance, they
may instruct a component to create a checkpoint and then stop once it
has reached a certain point in simulated time.  This naturally leads
on to the possibility of scripted steering and the ability to
``replay'' a previous, steered simulation.  We hope to investigate
these possibilities in future work.

\end{section}

%-------------------------------------------------------------------
\begin{section}{Summary}
\label{sec:Conclusions}

In this paper we have described a possible approach to the deployment
of coupled models on a general purpose Grid.  The coupled model
hybrid MD formed by molecular dynamics and fluctuating hydrodynamics
packages was illustrated for the case of a Couette flow
simulation. The two independent simulation codes communicate via a
HybridSwitch service and are launched via the RealityGrid launcher
application on different machines.  Computational steering and online
visualization were also described.  The problem of performance control is
particularly interesting for coupled models; we can provide this
initially via manual migration of both components between
computational resources on a Grid. In the future, automated
performance control may be included in the RealityGrid framework. The
generalization of the coupling protocol described here to the case of
multiple components will also be explored.
%Finally, we note that an essential requirement for future  deployment of large
%scale coupled models on grids is advanced reservation of co-scheduled resources
%since the various components have to be able to run concurrently; indeed,
%depending on the nature of the application, it may also be necessary to
%co-schedule an allocatable guaranteed quality of service network. Until now,
%even basic capabilities have been in short supply on today's production grids
%although efforts are underway to provide automated advanced, co-scheduled
%reservations of both resources and networks. 

\end{section}

\begin{acknowledgments}
This research was supported by the EPSRC RealityGrid
project GR/R67699 and the EPSRC Integrative Biology project GR/S72023.
\end{acknowledgments}

\bibliographystyle{unsrt}
\bibliography{realitygrid}

\end{document}